\RequirePackage{lineno}
\documentclass[12pt]{article} 
\usepackage[left=1.0in,right=1.0in,bottom=1.0in,top=0.6in]{geometry}                
\usepackage{fancyhdr}
\usepackage{setspace}
\usepackage{graphicx}
\usepackage{amssymb}
\usepackage{authblk}
\usepackage{multirow}
\usepackage{indentfirst}
\usepackage[superscript]{cite}

\fancypagestyle{styleone}{
\fancyhf{}
\fancyhead[L]{}
\fancyhead[R] {\fontsize{10pt}{12} \selectfont Proceedings of the International School of Physics ``Enrico Fermi" Course CLXXXVII \\Water: Fundamentals as the basis for Understanding the Environment and Promoting Technology \\Opening lectures, Italian Physical Society, Varenna, Italy, July 2013}
}
\title{Water and its relatives: the stable, supercooled and particularly the stretched, regimes}
\author[1,2] {S. L. Meadley\thanks{Stacey Meadley is a student volunteer who assisted in the transcription/condensation of the lectures of the senior author at the Varenna School course CLXXXVII.  The opinions/interpretations expressed in the manuscript are those of the senior author.}}
\author[3] {C. A. Angell}

\affil[1] {Department of Chemical and Biomolecular Engineering, Cornell University, Ithaca, NY 14850}
\affil[2] {Institut Lumi\`ere Mati\`ere, Universit\'e Claude Bernard Lyon 1, Villeurbanne, France, 69622}
\affil[3] {Department of Chemistry and Biochemistry, Arizona State University, Tempe, AZ 85287}
 
\begin{document}
\maketitle
\thispagestyle{styleone}
\begin{abstract}
While the water molecule is simple, its condensed phase liquid behavior is so complex that no consensus description has emerged despite three centuries of effort. Here we identify features of its behavior that are the most peculiar, hence suggest ways forward. We examine the properties of water at the boundaries of common experience, including stable states at high pressure, the supercooled state at normal and elevated pressure, and the stretched (``negative pressure") state out to the limits of mechanical stability. 

The familiar anomalies at moderate pressures (viscosity and density (TMD) behavior, etc.), are not explained by H-bond breaking, according to common bond-breaking criteria. A comparison of data on the TMD, at both positive and negative pressures, with the predictions of popular pair potential models, shows dramatic discrepancies appearing in the stretched liquid domain. This prompts questions on the second critical point (TC2) hypothesis that has been guiding much current thinking. We turn to related systems for guidance, reviewing a hierarchy of water-like anomalies. We conclude that water models are far from complete and that proper understanding of water will depend on success in mastering the measurement of liquid behavior in the negative pressure domain - which we discuss. 
\end{abstract}
\newpage

\section{Introduction}
Water is necessary for life on earth, providing the matrix for sub-cellular machinery to the vast ecosystems that exist in the world's oceans. Its presence in nature, the ways in which it has shaped the environment, how it acts as the driving force for atmospheric phenomena and climate, all make it arguably the most important of all compounds for us to study and understand. Beyond these considerations, water is a fascinating compound. The water molecule is formed simply of one oxygen and two hydrogen atoms producing a fundamental unit capable of linking to four of its nearest neighbors by hydrogen bonds (those largely electrostatic interactions between the protons on one molecule and the lone-pair negative sites on the oxygens of its neighbor molecules). The H-bonds are responsible for the existence of water as a liquid under most ambient conditions, see the phase diagram, Figure 1, and also the reasons that it is unusual as a liquid - though not completely unique as will see. 

Although water exhibits all the usual phase transitions characteristic of other substances, melting points, boiling point and gas-liquid critical point etc., it differs greatly in details such as the slope of the melting vs pressure line, (well understood  in terms of the negative volume of melting) and a much less well understood maximum in density (at 4$^\circ$C for ambient pressure). The temperature of the density maximum is a strong function of pressure, crossing the melting line at about 50 MPa and becoming undetectable below -25$^\circ$C and 120 MPa (for D$_2$O). The maximum becomes sharper with decreasing pressure and shows no sign of broadening near zero pressure.

\begin{figure}[ht]
\centering
\includegraphics[scale=0.12]{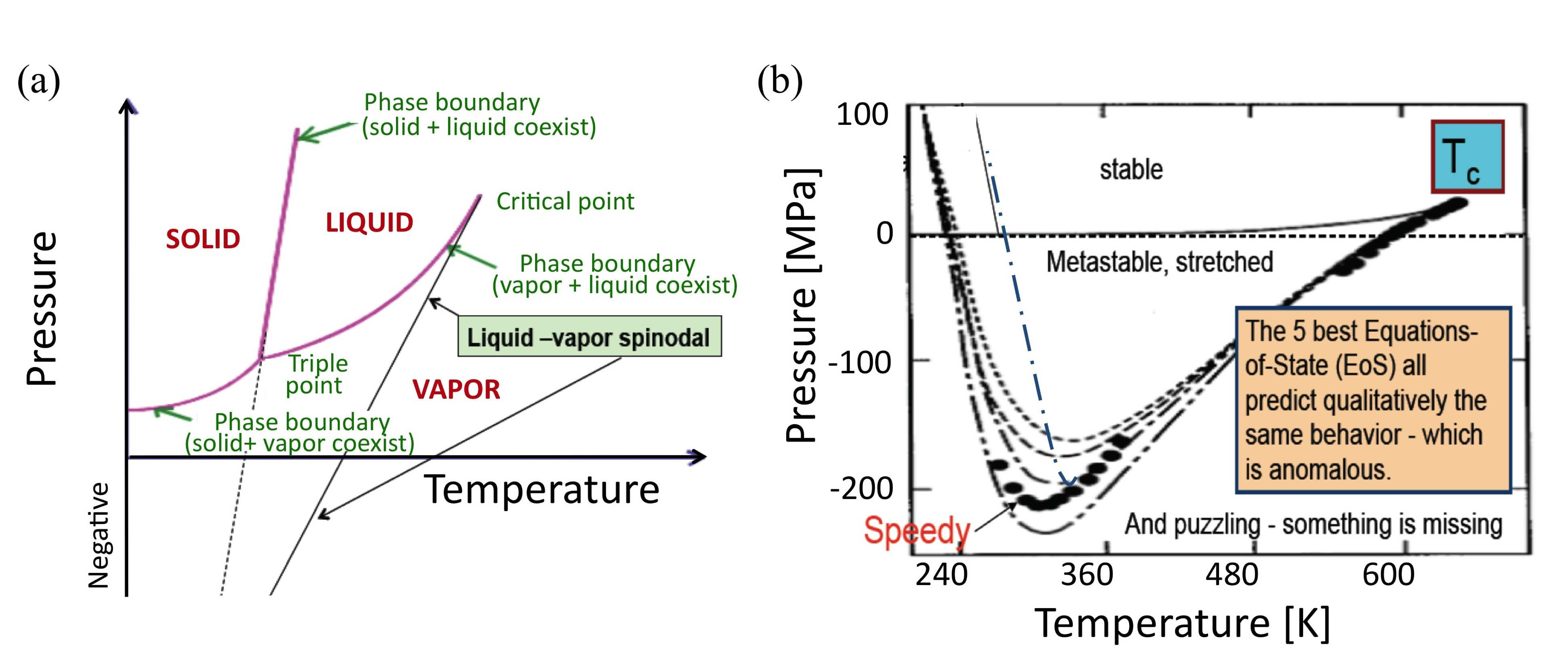}
\caption{(a) Phase digram of water seen in typical textbook, with addition of spinodal limit to metastable liquid states, including stretched states at negative pressure. (b) Stable and metastable liquid components of the phase diagram for water showing how all five major empirical equations of state show the liquid-vapor spinodal limit exhibiting an anomalous reversal of direction at approximately the point of intersection with an extrapolation of the line of density maxima (dash-dotted line) measured in the stable and metastable domains. This behavior is discussed in detail in the latter sections of this article.}
\end{figure}

As water's anomalies have been catalogued, concern over its behavior at temperatures well below 0$^\circ$C, i.e. in the metastable supercooled state, has come to the forefront. The expansivity has been known down to -33$^\circ$C since 1968 \cite{zheleznyi1969density} and more recently the heat capacity\cite{rasmussen1973,angell1973anomalous,angell1982heat} and compressibility\cite{speedy1976,kanno1979water} have been obtained. All appear to diverge at -45$^\circ$C in the supercooled liquid state.  An excellent over-view is provided by Debenedetti \cite{debenedetti2003supercooled}. 

What exactly happens to these properties below -38$^\circ$C is uncertain as the crystallization barrier blocks access to liquid water and we encounter what has been termed ``no man's land". In this contribution we address what we know and what we don't about water in these liquid states that is so different from the other liquids of common experience. Starting with the stable states we examine signatures of the -OH hydrogen bond network for water at high pressure. The exploration then moves to supercooled water and observations on aqueous solutions. Further, we ask whether water is unique or are there related substances that can help  us understand more about its structure and properties. Finally we reach the region of stretched water that has received much less attention. The stretched region, depicted in Figure 1b, is where the liquid is under tension (negative pressure). Thermodynamically water under tension is metastable but under the right conditions of preparation, it can exist in a mechanically stable state that can reach extraordinary levels of negative pressure, perhaps nearing -150 MPa, before cavitation to the vapor phase occurs. It is in this stretched region that we suspect the secrets of water's  anomalies will finally be unlocked and accordingly where we think experimental studies of water should be strongly encouraged, notwithstanding their evident difficulties.

\section{Liquid domains and stable states}
\subsection{Hydrogen bonds and anomalies of water}
H-bonds have been mentioned as the most important factor in describing water's uniqueness. But many liquids are multiply hydrogen bonded without showing any of water's anomalies. For instance, hydrazine N$_2$H$_4$ with two protons for each nitrogen lone pair, and H$_2$O$_2$,with two lone pairs for each proton, each have about  the same melting point as water, but have none of the anomalies - so it is more than just bonding that is involved. A strong clue is provided by one of the best known properties of water, one that is manifested by any iceberg in the ocean or, alternatively, any ice cube in a glass of water. We take it for granted that the ice is floating without realizing how uncommon this is amongst liquids. Compare this picture with liquid ethanol containing solid cubes of ethanol, which immediately sink to the bottom of their container. Normal liquids, like ethanol, have a lower density than their crystal forms, but liquid water has a higher density than crystalline ice and in fact exhibits a density maximum at 4$^\circ$C.

This comes about from the way that water ``sees" its neighbors and how the  H-bonds form between the two oxygen lone pairs and the two hydrogen atoms of each molecule. In the liquid state these bonds are in constant flux and the molecular network itself is in a state of constant collapse and reconstruction - whereas in the solid state - ice - the molecules choose to form an ordered hexagonal crystal with strictly four nearest neighbours per molecule and 180$^\circ$ O-H-O angles that require  a very open arrangement of the molecules - which guarantees low density. It proves critically important that, in water at low temperature, the thermal smearing effect that generates so much entropy at higher temperature, weakens, allowing the liquid to approach, more and more closely, the open network configuration of the crystal. This expansion overwhelms the normal positive vibrational contribution to the expansion coefficient, promoting the density maximum (famous for centuries) and then, at lower temperatures, the diverging negative expansivity reported by Zheleznyi.

Of course there have been a number of models developed to explain this behavior, and new ones are appearing regularly. They range from two-state models with only intact bonding or broken bonds, to entirely continuum type models. We do not attempt to review them here but rather will briefly examine some of the tools that are available to help understand the structures and restructuring that might be involved.

How can we observe these structural changes, in the laboratory? Methods that reflect the hydrogen bonding structure directly would seem the most appropriate. Vibrational spectroscopy, (particularly Raman scattering and infra-red absorption), has long been used for this purpose to look at the -OH oscillations in water. Mostly these have focused on the effect of temperature and pressure on the spectra\cite{walrafen1967raman,senior1969spectroscopic,luck1970approximate}. A new twist was added by Dana Fields who recorded changes in the spectroscopic signature of -OH as they were  liberated (from the H-bonded state) by chemical means\cite{angell1985spectroscopic}. The lone pair-hydrogen groups that exist in stoichiometric balance in water were disrupted by the addition of competing -NH groups of hydrazine, ammonia and methylamine. The resulting -OH spectra are strikingly similar to those seen, and associated  with ``freeing" -OH oscillators from their would-be neighbors, by increases in temperature, for instance in H$_2$O-D$_2$O solutions by Luck and Ditter\cite{luck1970approximate} and more recently down to -33$^\circ$C by Angell and Rodgers\cite{angell1984near}. The latter are shown in Figure 2a. In both cases the spectrum shifts across a centroid but two separate states cannot be distinguished and the bonded -OH character does not fully disappear. Results combining Raman spectra and computer simulations with SPC-E model suggest that  a continuum model as most appropriate for -OH bonding\cite{smith2005unified} though the SPC-E model fails badly to describe the behavior in the supercooled state. Controversies exist over whether the shifting of the spectra across a centroid is definitive of a weak-strong bond transition, but it is clear that a two-state model of water is insufficient.

\begin{figure}[ht]
\centering
\includegraphics[scale=0.12]{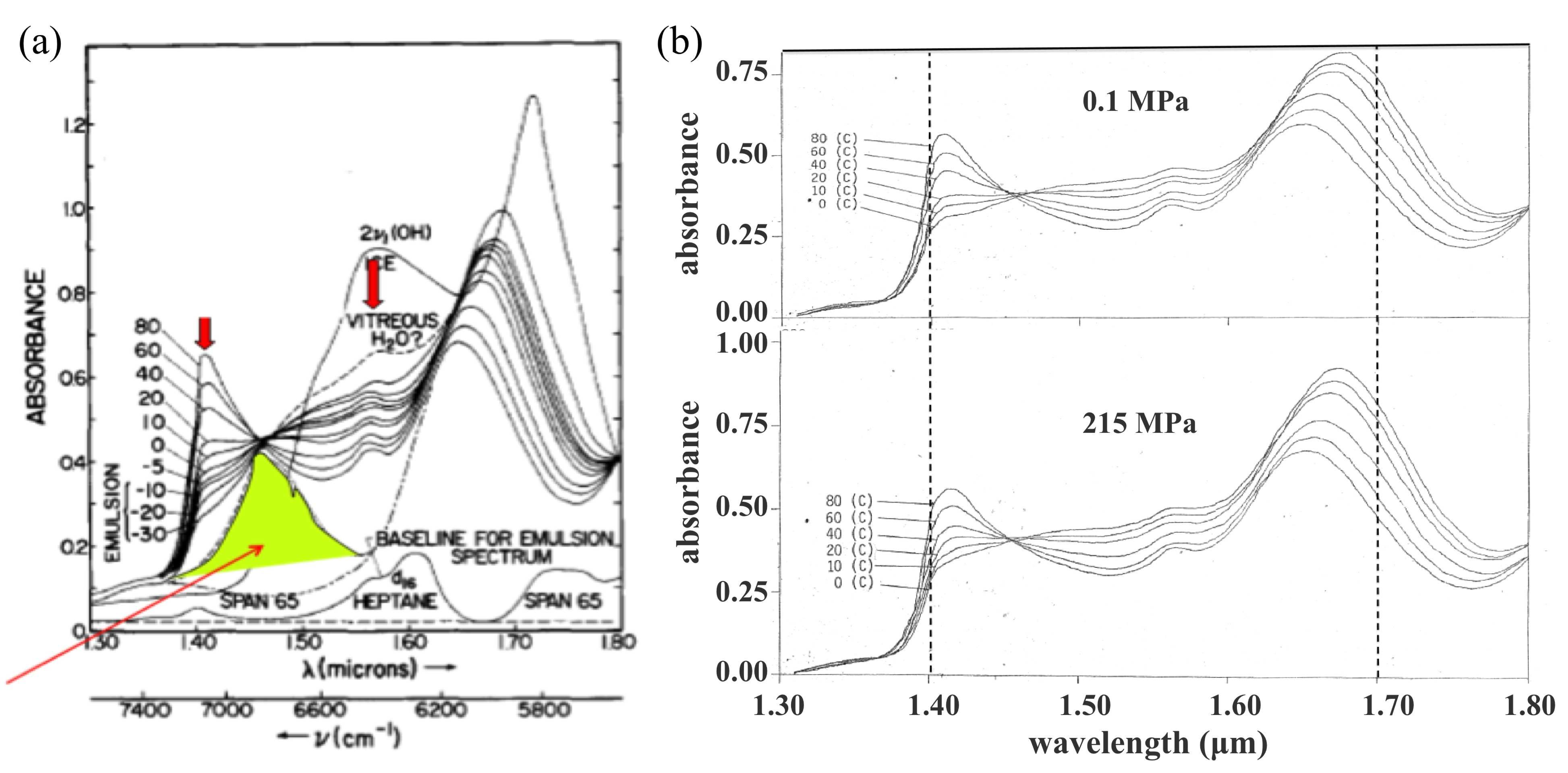}
\caption{Uncoupled overtone IR spectra of  H$_2$O in D$_2$O (a) in emulsion samples to enable spectroscopy over wide temperature ranges reaching -30$^\circ$C. Green-shaded area (color online) is left unaccounted for by two-state analysis  (b) as a function of pressure in sapphire window cell to permit study to 215 MPa in the temperature range 0-80$^\circ$C. No evidence for breaking of hydrogen bonds since free -OH band at 1420 mm is not enhanced by pressure increase. Figures adapted from ref \cite{angell1984near} and V.Rodgers Ph. D. Thesis , Purdue University 1982.}
\end{figure}

\subsection{High Pressure}
A key anomaly in water occurs in the high pressure region. It was observed long ago that increasing pressure causes viscosity to decrease, provided the temperature is low\cite{bett1965effect}. In ref [5], it was found there is a viscosity minimum as pressure increases, for any temperature below 33$^\circ$C. Additionally, the density maximum of liquid water is found to shift to lower temperature as pressure is increased. It is found that there is a correspondence of the viscosity anomaly and the temperature of maximum density (TMD) at high pressure. The density anomaly smears out at increasingly high pressure and, in D$_2$O, cannot be resolved below -25$^\circ$C. Simulations of the density anomaly in a range of models have been unable to reproduce the sharpness of the maximum. What is observed in some simulation models, for instance TIP4P, is that the density anomaly precedes a liquid-liquid transition that is implied, by extrapolation, to occur approximately 50K below the TMD\cite{molinero2008water}.

In laboratory water, the TMD changes more rapidly with pressure than the melting point so that, above 500 bar it becomes a supercooled property\cite{kanno1980volumetric}. It would seem natural to suppose that this is due to the effect of pressure breaking H-bonds. However, inspection of the IR overtone spectra that are supposed to reveal broken bonds by increase in IR spectral intensity of the ``free" -OH band at 1.4$\mu$m, show nothing happening. Comparing the ambient and 215 MPa spectra in Figure 2b one would have to say that there is no change in the structure of water despite passing into the region of high pressure where the density maximum and viscosity minimum have disappeared. Thus we encounter a classic case of ``what we know and what we don't"\cite{angell2001water}. We know anomalies exist but we have not been able to demonstrate what is the relation to the structure of water at least from the perspective of vibrational spectroscopy. Perhaps we need a spectroscopy that is more sensitive to the O-H-O bond angle, like NMR\cite{angell1973anomalous}.

Turning to simulations of water, the TIP4P model, when results are shifted to allow for the known displacements of simulated property from actual property in the familiar positive pressure range, is able to well reproduce the pressure dependence of the TMD\cite{corradini2010route}. The results of experiment, and of simulation, for the line of density maxima (LDM) are plotted in Figure 3 along with the TMD for real water according to the currently best regarded equation of state, viz., that known as IAP-WS95\cite{wagner2002iapws}. Representative points from the more extensive TMD studies available for D$_2$O\cite{kanno1980volumetric} are included after subtraction of 7K, and the TMD according to Speedy's simple EoS\cite{speedy1982stability} is added as a dashed line ending at the apex of his derived reentrant spinodal. 

While both the shifted TIP4P model and the IAP-WS95 EoS agree with the experimental TMD at positive pressure up to 100 MPa, the exact opposite applies to the relations at negative pressure. The shifted TIP4P model predicts a maximum in the TMD at about -120 MPa and 292K followed by a return to lower temperature, while the IAP-WS95 EoS yields values that turn in the opposite direction (red solid line). The latter locus is compatible with the temperatures where direct observations imply existence of a density maximum (two large solid circles), as will be discussed in our final section. Evidently the negative pressure domain offers new opportunities to discriminate between alternative models for water, as we explore in some detail, later.

\begin{figure}[ht]
\centering
\includegraphics[scale=0.1]{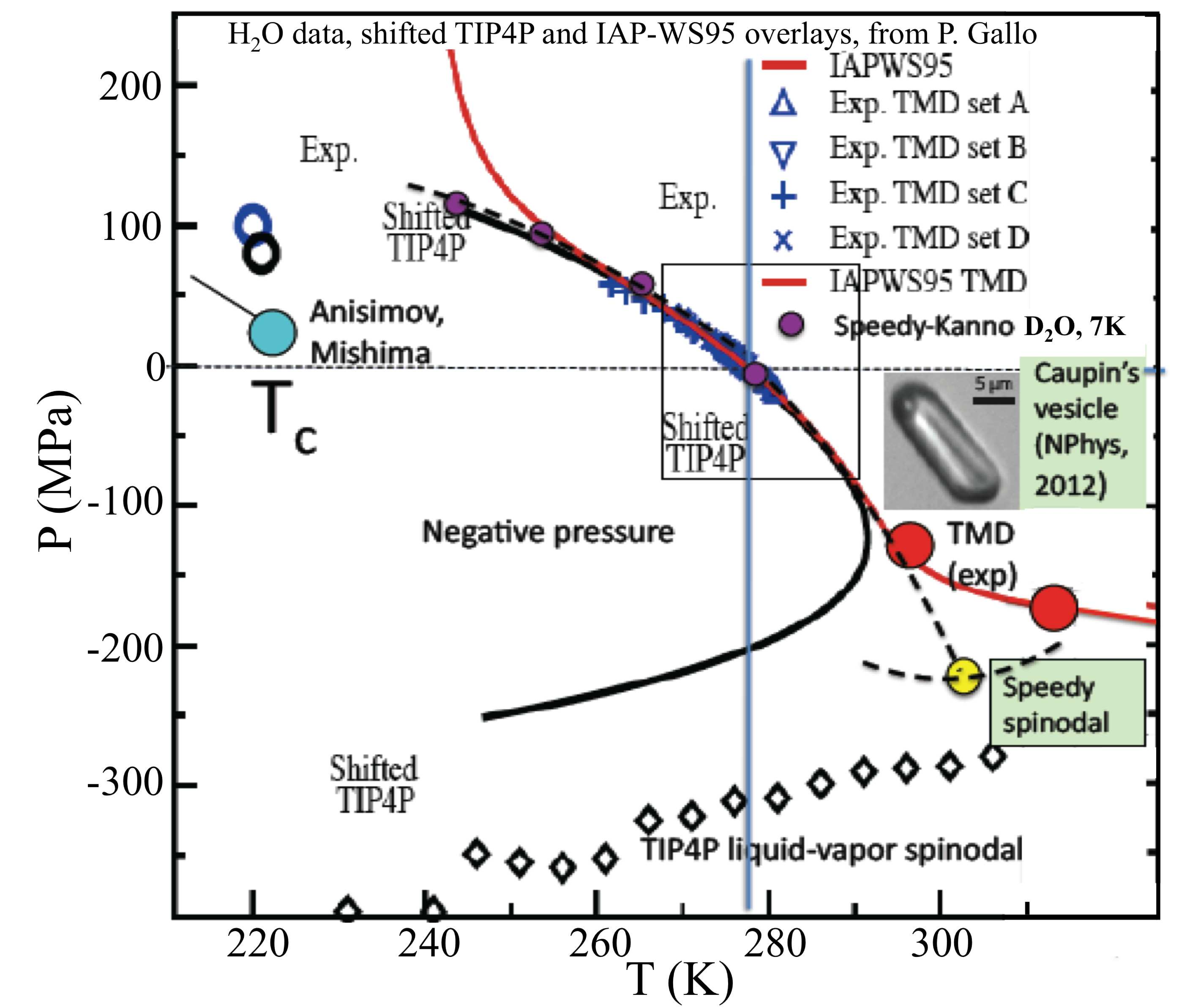}
\caption{Line of density maxima for TIP4P simulation model and IAP-WS95 equation of state showing good agreement with experiment at \textbf{positive} pressures (after a shift of the TIP4P findings) but divergent predictions at negative pressures, (figure adapted from a diagram by Sergey Buldyrev and Paola Gallo who superimposed the IAP-WS95 locus of TMDs on the figure from Ref.\cite{corradini2010route}. The TMD locus was calculated using the IAP-WS95 program by Steven L. Moshier - see Acknowledgements). The two solid circle symbols represent experimental points from different studies under tension, that will be discussed in a later section. The open yellow circle is the point of departure for the Speedy line of TMDs. Smaller solid circles are representative points for D2O from ref.\cite{kanno1980volumetric}, after downshifts of 7K. Insert: an example of the microscopic sample on which the negative pressure observations are made (from ref.\cite{azouzi2012coherent})}
\end{figure}

\section{Properties of supercooled water and solutions}
Recognising that the structural characteristics that lie behind water's anomalies must become better-defined at lower temperatures, many workers have delved into the realm of supercooled water. The existence of liquid water below the melting temperature is surprisingly widespread in nature. For instance, water droplets supercooled to -30$^\circ$C are ubiquitous in (high altitude) cirrus cloud formations, notwithstanding the fact that metastability is delicate and any stray nucleating particle can easily trigger crystallization. The stability of these droplets is due to the logarithmic relation between sample size and spontaneous nucleation probability. For this reason most of the successful studies of supercooled water have utilized samples of small dimensions. A brief history of studies of supercooled water is listed in Table 1. It is quickly seen that indeed the odd behavior found in the stable state becomes much more pronounced in the supercooled state.

\begin{table}
\caption{Timeline of developments in the study of supercooled water}
\begin{center}
\begin{tabular}[pos]{|l|l|p{8cm}|}
\hline
Date & Author & Experiment \\ \hline
1776 & Joseph Black & First review on supercooled water \\ \hline
1928 & Tammann, Schwarzkopf\cite{tammann1928anderung} & Determined the temperature of the density maximum down to -15$^\circ$C by varying pressure \\ \hline
\multirow{3}{*}{1951} & ``Rainmakers"\cite{mason1975clouds} & Cloud-chamber experiments are developed with the ability to supercool water down to -42$^\circ$C \\
& Wylie\cite{wylie1953freezing} & Arduously removed nucleating impurities in a 1mL sample of water until capable of supercooling to -31$^\circ$C \\
& Biggs\cite{biggs1953}, Mossop\cite{mossop1955freezing} & Supercooling of water in fine (2-10 $\mu$m) glass capillaries to -33$^\circ$C \\ \hline
1969 & Zheleznyi\cite{zheleznyi1969density} & Measurement of the density and expansivity in fine capillaries down to -33$^\circ$C \\ \hline
1970 & Angell \& Sare\cite{angell1970glass} & Studied glass formation in saline solutions and extrapolated to T$_g$ for water. (same value (136K) as from earlier molecular solution extraps). Demonstrated entropy paradox for deeply supercooled water. Raised possibility of, and defined, vitreous polymorphs.\\ \hline
1972 & Anismov, Voronel et al.\cite{anisimov1972specific} & Determined precise specific heats of  water down to -7$^\circ$C. From special curvature, surmised singularity at lower temperatures. \\ \hline
\multirow{2}{*}{1973} & Rasmussen et al.\cite{rasmussen1973} & Measured the heat capacity of supercooled water in emulsions down to -35$^\circ$C revealing large rise in C$_p$ as temperature decreases. \\
& Angell et al.\cite{angell1973anomalous} & Combined heat capacity measurements with NMR chemical shift data.  Proposed the analogy of the behavior of C$_p$ with magnetic transition in non-zero field. \\ \hline
\end{tabular}
\end{center}
\end{table}

In the 1970s, studies focused on the heat capacity (C$_p$) of supercooled water. The heat capacity in the stable state was already known to be high and to diverge upon heating into the superheated regime. Simple equations of state like that of van der Waals are able to predict this behavior. What would happen in supercooled states had not been examined below -10$^\circ$C until Rasmussen et al. were able to use emulsion techniques to examine micron-sized droplets of bulk water\cite{rasmussen1973}. They observed the heat capacity to increase rapidly as the temperature is cooled, particularly below -15$^\circ$C, creating a large difference in the heat capacity of supercooled water and ice. However below -38$^\circ$C crystallization could no longer be avoided. At ambient pressure the increase in heat capacity for supercooled water at -38$^\circ$C is the same as the increase observed for water at its superheating limit\cite{angell1982heat}. In the superheated case the behavior of the C$_p$ is known to be related to the spinodal and corresponding liquid-vapor critical point, which revived the earlier speculation\cite{anisimov1972specific} that the supercooled and superheated water behavior might share common (pre-spinodal) origins. Counteractively, however, the validity of the  measurements themselves was questioned, with the suggestion that the anomalous heat capacities being reported were an artifact of the small sample (microdroplet) strategies employed in their measurement\cite{johari1997gibbs}. However, this popular criticism was effectively put to rest by the demonstration that heat capacities measured in 10ml ampoules of pharmaceutical grade water, down to -30$^\circ$C by Tombari et al.\cite{tombari1999heat}, agreed perfectly with the previous emulsion results (see the recent review by Holten and Anisimov of all previous work on this issue\cite{holten2012thermodynamics}). 

Since there is no longer any doubt about the existence of these anomalous response functions, and their accord with critical divergence functions (after a ``normal" component is subtracted from the total), nor about the anomalous power law variations of the various transport measurements that have been made\footnote[1]{The power laws were established on the basis of data obtained using all manner of techniques (viscosity to -35$^\circ$C, dilute solution electrolyte conductivities, diffusivities dielectric relaxation times, NMR reorientation times to -38$^\circ$C), all of which were fitted by the same equation $X=A_x(T/T_s-1)^{-\gamma}$ with the same divergence temperature of 228K.}, it remains to provide an explanation for their origin. For some time there had been a consensus of sorts  that this explanation had been provided by Poole et al.\cite{poole1992phase} (Stanley group at Boston University), but recently this has become a matter of major controversy. We will deal with this in the next section.

\section{The second critical point hypothesis and the problem of its location}
Until the paper by Poole et al.\cite{poole1992phase} , which described computer simulation studies of a particular pair potential model for water (the ST2 potential) proposed explanations for the alarming anomalies had been sparse. An early suggestion due to Rasmussen and McKenzie invoked Frenkel fluctuations anticipating a crystallization event\cite{rasmussen1973clustering}. This notion was recently revived by Limmer and Chandler\cite{limmer2011putative} without details, in their criticism of the concept of liquid-liquid transitions and second critical points in water and other tetrahedral liquids . Then there were two alternative proposals from Speedy and Angell, (elaborated in one case as the ``stability limit conjecture" by Speedy\cite{speedy1982stability}, and then, in 1996, a ``singularity-free proposal" by Sastry et al. in the Debenedetti group\cite{sastry1996singularity}. 

The stability limit conjecture assumed a fundamental cut-off for mechanical stability of the condensed phase for water that grew from the well-established spinodal limit for stability with respect to irreversible cavitation predicted  by the  van der Waals equation. For normal liquids, the spinodal limit curves monotonically away from the critical point (see Figure 1 (b)) to meet 0 K at large negative pressures. However, by data-fitting to a simple Taylor expansion about the conjectured spinodal, Speedy\cite{speedy1982stability} deduced that for water, the spinodal temperature, T$_s$, departed from the normal trajectory and returned to positive pressures.  Note that only the first two terms of the Taylor expansion were needed to fit the exquisitely precise measurements on water density in the temperature range 0-100$^\circ$C and pressure range 1-1000 bar.

The Speedy spinodal limits are reproduced in Figure 1(b) where they are seen to lie close to the spinodal boundaries predicted by the much more eleborate multi-parameter equations of state of the water and steam engineers. Indeed, Speedy's spinodal limits extrapolate back to positive pressures at a temperature very close to the temperature predicted\cite{kanno1979water} by power law fits to the divergent compressibility data of supercooled water -45$^\circ$C. Thus the spinodal terminated the free energy surface for water, not only at negative pressures but also at positive pressures. 
The alternative Speedy-Angell scenario was less dramatic; it was not given a quantitative basis and so remained undeveloped. It proposed\cite{speedy1976}, rather than a termination of the free energy surface, a quite abrupt step on the free energy surface, which these days would be described as a Widom line. At this line the response functions would pass through a sharp extremum as depicted in Ref. 3, a behavior that was interpreted by reference to magnetic phase transitions\cite{angell1973anomalous}. 

Finally, the singularity-free model, based on a lattice model with tetrahedral coordination preferences in the formulation, yielded sharp maxima in response functions which had their extrema at a different temperature for each function, and a phase transition located at 0K\cite{sastry1996singularity}.

The Poole study\cite{poole1992phase} began as a search, by computer simulation using the ST2 potential, for the Speedy spinodal, the extremum of which was expected to occur at negative pressures, where the temperature of maximum density intersected the liquid-vapor spinodal. They did not find such an intersection but instead found an anomaly in the pressure-density relation at positive pressures which, at the limit of simulation capability, developed into a van der Waals loop. Their paper announced the discovery of a second critical point in the case of liquid water, and began a new era in water studies by showing how the anomalies observed could be linked to critical point fluctuations. 

Two major problems emerge. The first concerns the validation of the hypothesis for models, and then for real water. The second, conditional on a positive resolutions of the first, is where, in temperature and pressure,  does the second critical point lie, for real water.

\subsection{Validity of the second critical point hypothesis}
\subsubsection{Water models}
The second critical point hypothesis was advanced on the basis of findings on a particular model of water, the ST2 model which has since been exhaustively studied. Notwithstanding the many self-consistent features of the findings, such as the obvious van der Waals loop that has been observed (without any interference from crystallization on the time scale of the calculations by Poole et al.\cite{poole1992phase}, see section 4), the validity of the findings has been challenged recently by Limmer and Chandler\cite{limmer2011putative,limmer2013putative}. The claim of these authors, made on the basis of free energy calculations using the \textit{hybrid Monte Carlo} calculation technique, is that there is only one well-defined free energy basin for water rather than the two that should be found anywhere within the spinodal boundaries below the 2nd critical point in the second critical point scenario. However this claim has been countered by free energy calculations from the Poole, Sciortino\cite{poole2013free} and particularly the Debenedetti\cite{liu2012liquid,palmer2013liquid} groups, who have used multiple methods and all obtained two distinct basins for the ST2 model\cite{holten2012entropy,holten2014two}. Of course, since the problem is set in the metastable state it is just a matter of coarse-graining the timescale sufficiently to ensure that only one basin will survive, and it is inevitable that the LDL phase will be the shortest-lived because it is the phase that is topologically closest to that of the crystal. The important point, however, is whether the LDL phase has a long enough existence to be observed in the internally equilibrated state. This seems to have been unambiguously established in the most recent work\cite{palmer2013liquid}. Resolution of the conflict will no doubt soon arrive.

Validation of the LDL phase for ST2 does not, in any case, validate the 2nd critical point for other models and certainly does not mean that real water has a second critical point. The occurrence of such a singularity depends  on a delicate balance of many factors. Each case must be validated separately. It is clear that in the case of the mW model of water (which lacks rotational degrees of freedom, and so is very fast to equilibrate)\cite{molinero2008water}, the crystallization of ice I$_c$ occurs so readily that there is no hope of observing any slow fluctuations associated with critical point behavior. If it can't be observed it doesn't really exist. Of course virtual transitions are given credibility as explanations of  interesting phenomena. The Landau de Gennes (LdG) singular point underlying the weak isotropic-nematic transition in liquid crystal phenomenology\cite{de1993physics} is an outstanding example, and was part of his Nobel prize recognition. The LdG singular point cannot be reached because of rapid transformation to the lower free energy phase during cooling. Likewise there are hidden critical points in binary systems that don't quite unmix, that explain liquid state anomalies. Always the preferred explanation will be the one that best accords with all the observables.
\subsubsection{Real water}
Evidently the  question of timescale for crystallization relative to the time scale for  observation of the liquid properties is the major problem blocking any verification of the critical point, or second liquid state, in real water\footnote[2]{This time scale difference can be manipulated to an extent, e.g. by changing conditions so that the system time scale changes. Glass scientists usually add second and third components, or occasionally change pressures, to lengthen the crystallization time scale. The former has been applied by Murata and Tanaka recently, using  glycerol to reveal a sort of spinodal decomposition that generates a new liquid phase\cite{murata2012liquid}. Using pressure to open the time scale gap has not been systematically exploited to date.}. Most of the evidence relating real water behavior to second critical point fluctuations has come from supercooled water studies conducted between the melting point and the homogeneous crystallization temperature at various positive pressures. The findings have been consistent with the existence of a second critical point without confirming its reality or establishing its T$_c$,P$_c$. Here and later we consider evidence, much of it still to be obtained, from other domains.

Unfortunately, no way has been found, even using the smallest purest samples, including some with only 5000 molecules\cite{bartell1994supercooling}, to avoid homogeneously nucleated crystallization during slow cooling. However, very rapid cooling (``hyperquenching" $\sim$ 106K/s) can bypass crystalline ice  and form glassy water\cite{angell2008insights} at a temperature that is far below any of the above-deduced special transition temperatures. For the discussion of amorphous ices we refer to Loerting \& Giovambattista\cite{loerting2006amorphous} and Mishima\cite{mishima1996relationship}. Here we only note that the glass formed by hyperquenching is essentially the same material, by any structural measure, as the glass originally formed by vapor deposition\cite{burton1935crystal}, or the glass formed more recently by high pressure crystal amorphization and annealing\cite{mishima1984melting,mishima1996relationship}.

The findings on glassy or amorphous water during warmup, by researchers looking for the usual glass transition, have been the source of enormous controversy extending over decades in time, the resolution of which can make an important contribution to the critical point discussion. As is normal, the heat capacity of the \textit{glassy} state is only slightly higher than that of crystalline ice. But at the temperature, 136K, of the glass transition that is indicated by nearly all extrapolations of binary solution data to pure water, the large jump in heat capacity predicted by the corresponding extrapolations of binary solution heat capacity jump data  ($\sim$ 20 J/mol.K)\cite{angell1980heat}, is nowhere to be seen. One study of the glass formed by vapor deposition (at rates ten times slower than the minimum for 100$\%$ vitrification identified by Olander and Rice\cite{olander1972preparation}) directly into a DSC pan, subsequently scanned at the usual 20K/min using the reliable Perkin Elmer DSC-4 calorimeter, yielded an outright null result\cite{macfarlane1984glass}. Others, using various pre-transition annealing procedures known to enhance the magnitude of glass transition warmup effects, have found heat capacity increases of marginal  detectability with magnitudes in the range 0.7-1.6 J/mol.K\cite{hallbrucker1989glass,amann2013water} (only 1/15th or less of the value predicted by the binary solution extrapolations). The result is particularly vexing in light of the abnormally large jump in heat capacity, at the same temperature, exhibited by fully hydrogen bonded mixtures of water with water-like second components, hydrazine N$_2$H$_4$, and hydrogen peroxide H$_2$O$_2$\cite{oguni1980heat,angell2008insights}. Clearly there is something special about the heat capacity function for pure water.

In light of recent definitive measurements, with supporting evidence from relaxation time data\cite{amann2013water}, there can no longer be much doubt that the transition at 136K is the thermal identity of the ergodicity-restoring transition, i.e. the glass transition, but it is a transition from glass to a liquid that has the most ideally ``strong" liquid behavior on  record. Since the liquid state that has been studied by supercooling between 0$^\circ$C and -38$^\circ$C (discussd in the previous section) is arguably the most ``fragile" liquid known, it is clear that some major structural event occurs in water as a function of both temperature change (between 235K and 150K), and of salt concentration between 0 and 10N at constant low temperatures near 136K\cite{ito1999}. Our next task is to explain what sort of event that might be, and how it might reasonably connect the tiny excess heat capacity of the low temperature liquid to the extraordinarily large heat capacity of the supercooled liquid at -38$^\circ$C . This was the focus of a recent major article by one of us\cite{angell2008insights}.

While the behavior of C$_p$ in ``no man's land" could not, until recent ultra fast scan microcalorimetry studies,\cite{bhattacharya2011bulk,sepulveda2012glass} be determined directly, the laws of thermodynamics place restrictions on how it might change under the assumption that the change is continuous. For instance, the observation that C$_p$ of supercooled water rapidly rises before crystallization but is near the ice value in the measured amorphous form at 136K, poses an entropy problem\cite{angell1973anomalous} as follows. If heat capacity were to continue to increase below -38$^\circ$C, all the excess entropy gained on fusion would be lost by -65$^\circ$C (the supercooling liquid is losing entropy much faster than the corresponding crystal)\cite{angell1973anomalous}, far above the 136 K glass transition. A detailed analysis was provided by Starr et al.\cite{starr2003prediction}, placing the limits on the forms it could take, but not excluding the most interesting one. The most interesting one, that could only coincidentally be the right one, is illustrated in Figure 4, third panel.  Figure 4 shows the behavior of two condensed phase systems with critical points of the Ising universality class, which is the class to which a second critical point in liquids would be expected to belong. These are frequently called $\lambda$ transitions, so-named because the shape of the heat capacity curve resembles the greek letter $\lambda$. This type of behavior is seen near critical points of the isooctane-perfluoroheptane binary (liquid) system\cite{oby2001heat}, and Zn-Cu\cite{ashman1969specific} and Fe-Co (solid) alloys\cite{sato1976kinetics} of which the first and last are shown in Figure 4.  In the case of water, the $\lambda$ transition would be expected to describe the heat capacity characterizing a second critical point that just happens to lie at ambient pressure. This would be an unimaginable coincidence, since there is no agreement on the location of the second critical point, simulations of ST2 and TIP4P-2005 water predicting it to occur at much higher pressures and other cases predicting that it lies at negative pressure or even vanishes into the spinodal.

\begin{figure}[ht]
\centering
\includegraphics[scale=0.15]{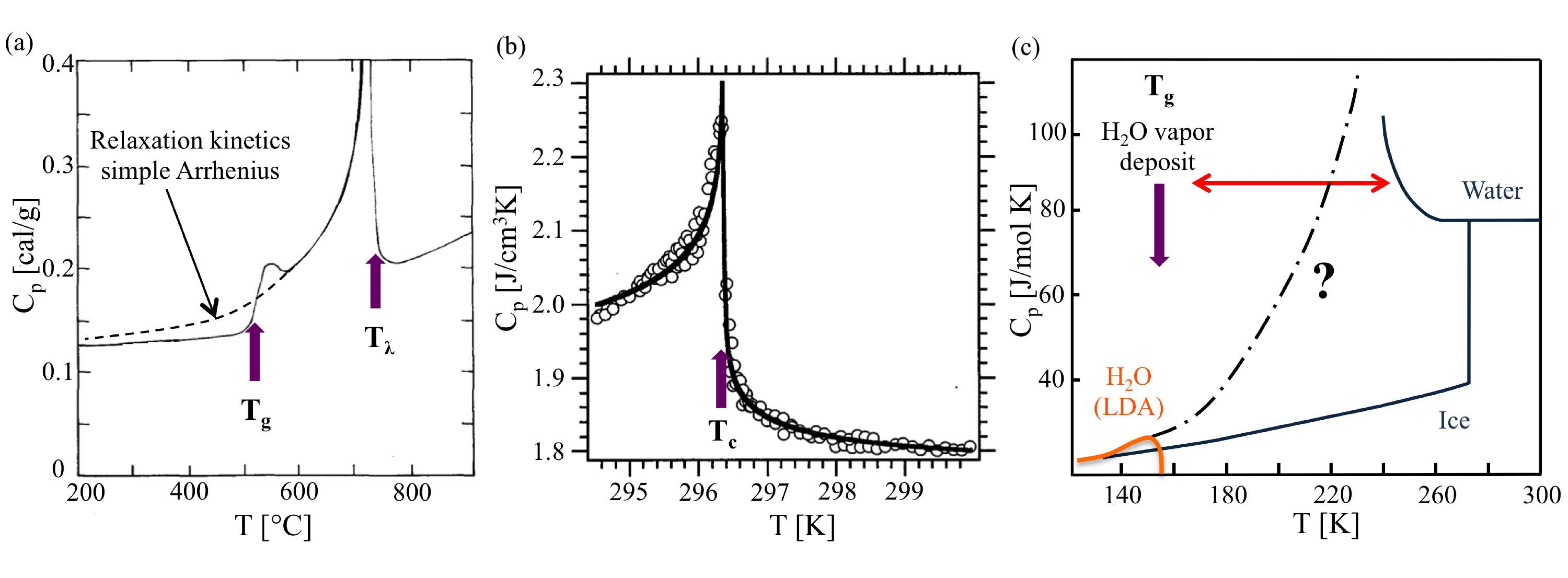}
\caption{Heat capacity forms for cooperatively disordering systems that generate  a critical point. From left, (a) solid Co-Fe ordered alloy\cite{sato1976kinetics} (two inter-penetrating simple cubic lattices, becoming random BCC above critical temperature Tl) (b) separate isoctane and perfluoroheptane-rich phases becoming a single phase solution with infinite clusters at the critical temperature\cite{oby2001heat} (consolute  point at 296.4 K) and (c) rationalization of the observed heat capacity behavior of water, on either side of ``No-man's land" (horizontal red bar) where only very short time scale measurements can be made, see text.}
\end{figure}

Figure 5 shows the possible behavior of \textit{the ambient pressure} C$_p$ for cases of a second critical point pressure different from ambient. If T$_c$ were to exist but lie at positive pressure, the ambient pressure C$_p$ would show a rounded peak (Figure 5a), a case now known as a Widom line crossing. The other case, showing ambient C$_p$  expected if T$_c$ lies at negative pressure, is illustrated in Figure 5b. In that case a first-order transition with substantial transition enthalpy, would be seen in an ambient pressure scan. This presentation of evidence may suggest that the second critical point scenario is the most probable, however, without measurements in ``no man's land" or further proof, there is insufficient reason to conclude this is the case and it remains an open question. 

\begin{figure}[ht]
\centering
\includegraphics[scale=0.15]{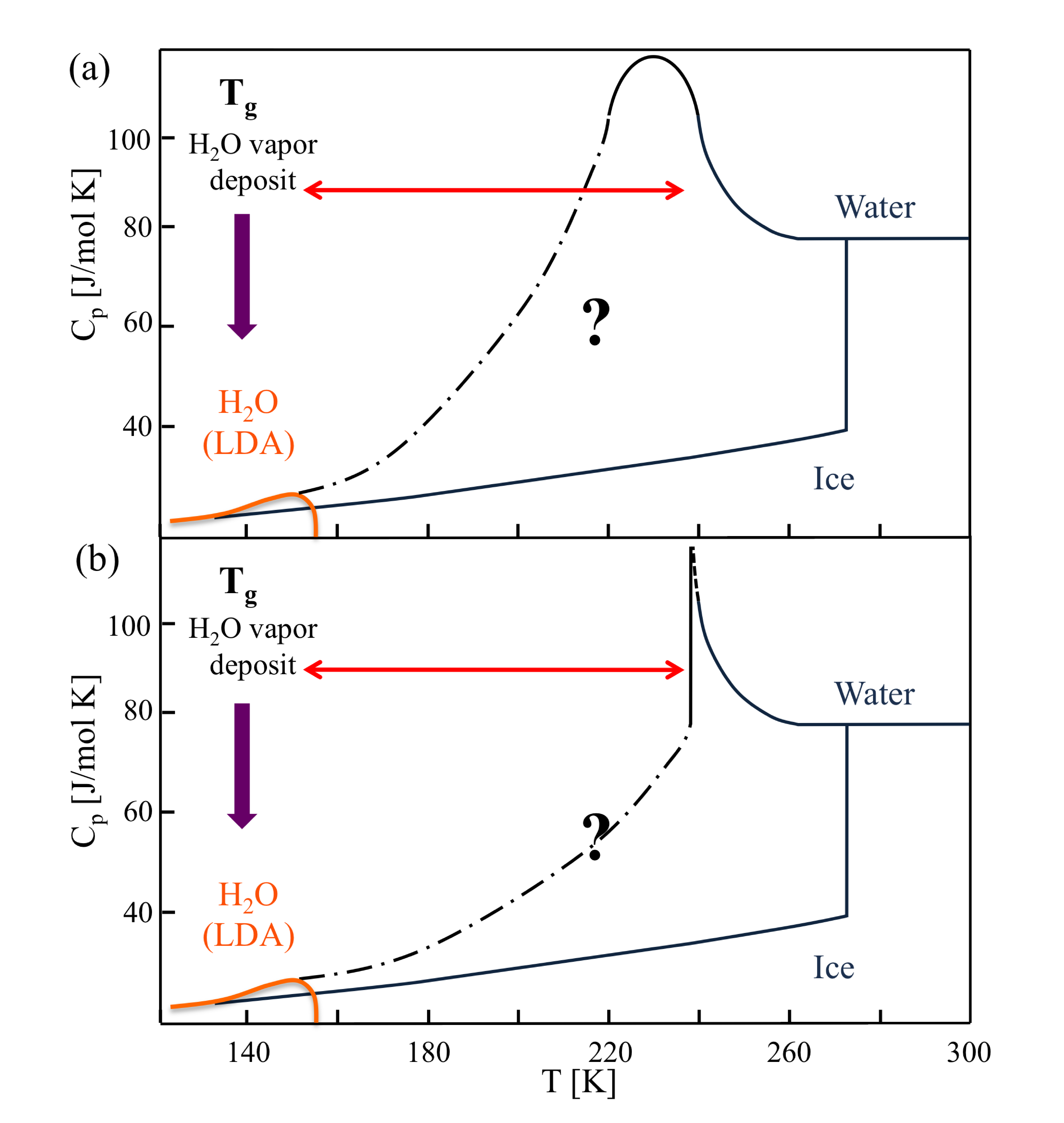}
\caption{Possible behavior of the ambient pressure  heat capacity of supercooled water in the presence of a liquid-liquid critical point (a) at positive pressure and (b) at negative pressure. Double headed arrows indicate the temperature range that is inaccessible to all except very short time measurements (popularly known as ``no-man's land").}
\end{figure}

The heat capacity is only one of several properties where anomalous behavior is amplified on supercooling. A summary of properties, including isothermal compressibility and viscosity, exhibiting power law divergences was presented by Speedy \& Angell\cite{speedy1976,angell1983supercooled} and more recent work on the volume of supercooled water is presented and discussed by Mishima\cite{mishima2010volume}. The behavior of these properties also suffers from the same limitation as the heat capacity insofar as we do not know what happens below the temperature of homogeneous crystallization. Reaching lower temperatures has been achieved only by going to higher pressures\cite{kanno1975}. Kanno showed the isothermal compressibility increases with supercooling for pressures up to 190MPa\cite{kanno1979water}. Kanno's extension of Speedy's ambient pressure study\cite{speedy1976} provided a ``nest" of compressibility vs. temperature relations each of which could be fitted with a power law divergence temperature (T$_s$) decreasing with increasing pressure. In fact, when the uncertainty of fitting was reduced by keeping the power law exponent constant, the divergence temperature closely tracked the line of homogeneous nucleation temperatures, T$_h$, known from previous work. The locus of T$_s$ values, which was interpreted as the spinodal limit to stability of high density phase of water, lay some 8-15K below the T$_h$ locus. Indeed, it appeared to be a continuation of the spinodal line bounding the negative pressure domain of Figure 1b.

Although some penetration into the forbidden zone below T$_h$ is currently being obtained by the streaming micro droplet x-ray laser studies of Nilsson and coworkers (A. Nilsson, contribution to this volume), we are generally prevented from direct observations of events near T$_s$, by pre-emptive crystallization. An alternative line of enquiry is the study of a small number of chemically very different liquids that share a surprising number of water's anomalies.

\section{Is water unique? Brothers \& cousins}
Having explored some of the peculiarities of water we need to ask whether water is truly unique or does it belong to a family of compounds?  If water is to have relatives, they will most likely belong to some group of tetrahedral structure forming species that also exhibit a density maximum. The brother of water would certainly be silicon, with germanium as the younger sibling. The extended family consists of cousins SiO$_2$ and GeO$_2$ and more distant relatives like Tellurium and Ge-Te alloys. Studies of these types of structured compounds seek to define the phase behavior and provide insight into our consideraton of water. 

Water's closest relative, silicon, features an increasing heat capacity as it is cooled below T$_m$ (1700K) similar to that seen with water. Molecular simulations of liquid silicon extending the range of cooling have shown that a sharp drop in heat capacity at the liquid-liquid transition temperature to a value that is barely in excess of the harmonic solid value\cite{sastry2003liquid}. This behavior is proof of ``strong" liquid character, where the viscosity increases in an Arrhenius-like fashion, while in the HDL phase at higher temperature it is a ``fragile" liquid. The similarity to Figure 5(b) is obvious. The diffusivity of silicon drops by 4 orders of magnitude across the transition.

Additionally, the simulations probe the short time dynamics above and below the transition temperature. Below the critical temperature slow dynamics dominate while above this temperature fast dynamics are seen. Fragile-to-strong transitions have been under discussion for water for two decades. Other similarities between water and silicon are the density maximum\cite{angell1996glass} and diffusivity maximum\cite{vasisht2011liquid} that shifts to higher temperatures as pressure is increased (as known also for water). Further evidence of a critical point is seen in the pressure-density isotherms. As is well known from the van der Waals equation of state, the pressure-density isotherms exhibit an inflection at the critical point and form a loop at lower temperatures. These are seen in the Figure 6 simulations of both ST2 water\cite{liu2012liquid} and the S-W model for silicon\cite{vasisht2011liquid}. A direct consequence of the isotherm behavior near a critical point, is that the \textit{isochores} in a P-T plot tend to cross at the critical point\cite{poole2005density}. Thus the isochore behavior approaching the critical point can serve as a pointer to the latter, if it exists.

\begin{figure}[ht]
\centering
\includegraphics[scale=0.1]{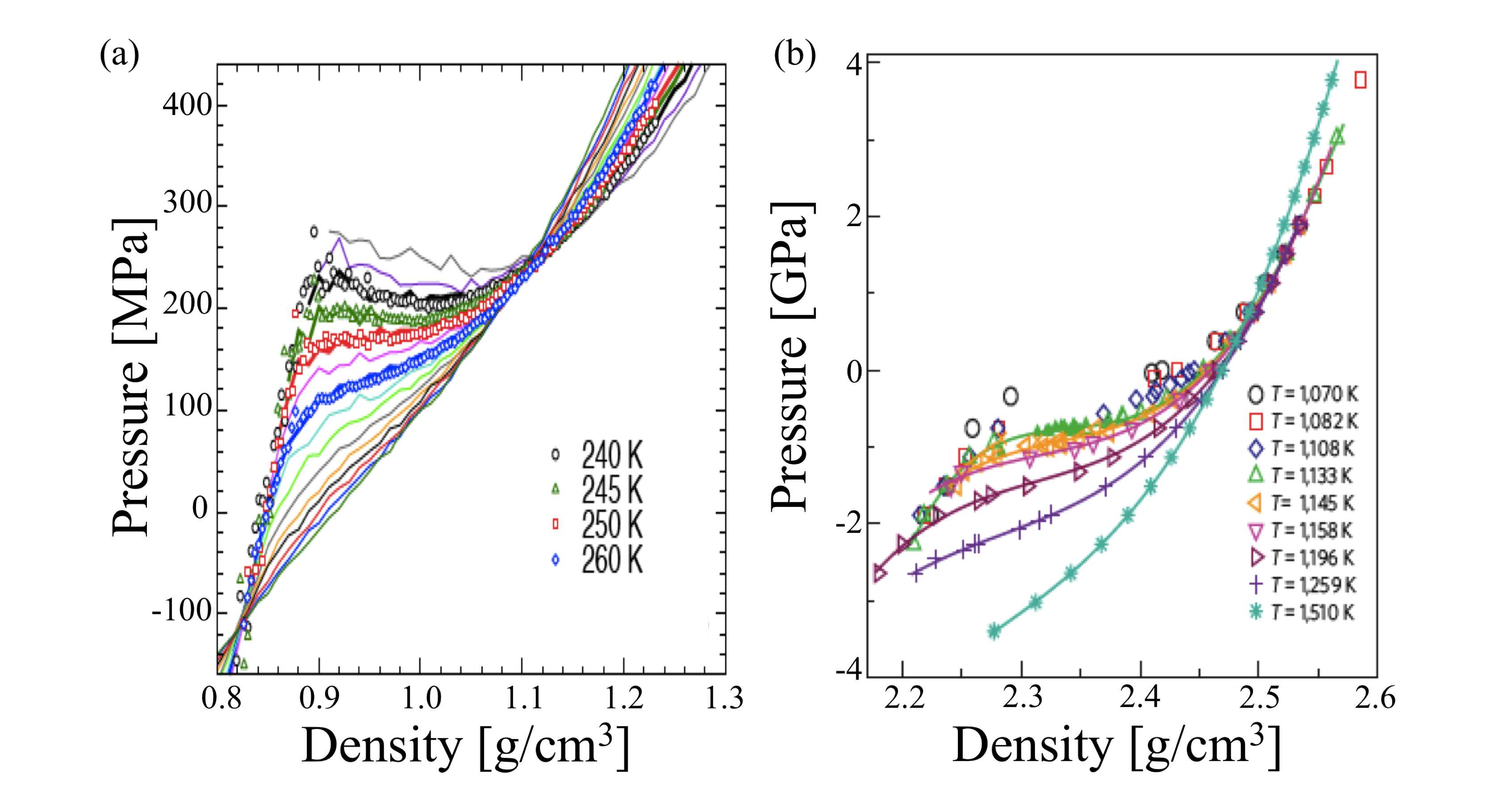}
\caption{Inflection behavior of isotherms in simulations of (a) ST2 water and (b) S-W model for silicon.}
\end{figure}

Before considering this behavior any further, we note that remarkable similarities have been observed between supercooled water and some more ``distant" relatives. Kanno showed there is a connection between the thermodynamic derivative properties of tellurium and water\cite{kanno2001}. After temperature scaling, the compressibility, expansivity and heat capacity almost exactly superimpose on the data for supercooled water at higher temperatues. However, where tellurium is able to capture the heat capacity maximum that is only assumed in Figure 4(b) (or predicted by some models) the water plots keep diverging. Other properties only seen for water in confinement or in ST2 at pressures lower than that at the critical point are captured in relatives like As$_2$Te$_3$ in which both a density maximum and a density minimum are clearly seen\cite{tver1996heat} - both of them above the melting point. Perhaps even more striking is the fragile-to-strong transition that is observed in many metallic glassformers\cite{zhang2010fragile} which can probably be related to the behavior of a simple monatomic liquid model, the attractive Jagla model\cite{gibson2006metastable,xu2009monatomic} in which a liquid-liquid critical point is known to occur above the melting point. This behavior, in the metallic glass material class that would not be considered to be water-related, demonstrates that there are many discoveries to be made in this field. 

There are certainly a number of parallels to be drawn between water and its fellow tetrahedral compounds. Although water has a unique place by virtue of its abundance and importance to life, it should obviously be considered as part of this tetrahedral family and a member of an even larger groups of compounds, that includes the  bulk metallic glassformers. While it is encouraging to see the anomalies predicted for water exhibited in its related compounds, we are still unable to determine what is the exact behavior of water due to the crystallization barrier. 

\section{Stretched water: negative pressure investigations}
Having explored the available liquid forms of water and its related tetrahedral structure-forming compounds there remains the possibility that the answers to water's strangeness may lie in the negative pressure or stretched water regime. We saw a hint of this in Figure 3, where the interesting part of the line of density maxima (LDM) seemed to lie in the negative pressure domain. 

The shape of the LDM according to the IAP-WS95 EoS and the TIP4P simulations, which accord well at positive pressures, branch in almost opposite directions as they descend into the negative pressure region. If the IAP-WS95 model were truly to represent the behavior of water, we must then expect the LDM to merge with the liquid-vapor spinodal which would then have to exhibit the retracing behavior\cite{speedy1982stability} of the original Speedy conjecture. Should the TIP4P scenario be correct, the LDM would pass through a maximum and retrace at lower temperatures while the spinodal would continue in a monotonic fashion (the two would not intersect). The latter case would be indicative of a second liquid-liquid critical point at positive pressures, as seen in Figure 3 for TIP4P and as has been seen in a number of simulations of water models\cite{liu2012liquid,abascal2010widom} or weakly negative pressures as has been suggested less frequently.

The former case clearly matches the best expectations of a succession of seasoned empiricists working with the best water data base available. The usual criticism of authors who take the EoS of empirical water analysts (i.e. Figure 1b) seriously, is the solid observation that no multiparameter equation can be trusted beyond the range of its data base. However such critics then need to account for the similarity of behavior derived by Speedy who found that he needed no more than two of his Taylor expansion coefficients to fit the high precision density data in the range of his analysis (0-100$^\circ$C) within their precision limits. One of the free parameters in his simple EoS, 
\newline
\indent $1-p/p_s = B(V_s/V-1)^2$
\newline
\indent \textit{where} $B=-(\partial ^2p/\partial \rho ^2)_{T,p_s} / 2p_sV_s$ , \textit{p and $\rho$ are the experimental density and pressure,
\newline
\indent and P$_s$ and V$_s$ depend on the chosen temperature},
\newline
is the pressure of the stability limit P$_s$ for each isotherm. The corresponding  spinodal limits P$_s$(T) are shown as large black dots in Figure 1b (also shown as a dashed curve in Figure 3). The Speedy spinodal lies within the envelope of the others in Figure 1b. It has a tensile extremum that is 40MPa more negative than that of the IAP-WS95 EoS which itself lies 20MPa more negative than that of the HGK (former National Bureau of Standards) EoS.

These considerations provide a compelling motivation for experimental and theoretical exploration of the negative pressure regime, which has been largely neglected in the long history of water and aqueous solution studies. The reasons for neglect are not difficult to find. The concept of negative pressure is foreign to most workers and, more important, measurements under large tension are difficult to perform. Furthermore, they seem to be blocked by a new sort of ``no-man's land" to be discussed in our final section that deals with such experimental data as have been obtained to date.

Before that, however, it is important to emphasize what is missing from the empirical EoS and Speedy stability limit scenarios. What is missing, and seriously troubling, is any explanation of what state of water is represented at temperatures below the retracing spinodal. This deficiency is now rectified by the recognition that the spinodal, in the zone of retracement, must be a spinodal \textit{on the low temperature side of a liquid-liquid coexistence line}, which lies within the ``no-man's land" occupied by the highly metastable LDL domain. This spinodal is the limit of mechanical stability for ordinary water (now referred  to as \textit{the HDL phase}), as in Speedy's conjecture. i.e. the line of temperatures at which any fluctuation will grow without restoring force to take the system, by first order non-equilibrium phase change, out of the HDL phase. However, the phase to which the water transforms is now recognized as \textit{the LDL phase}.

Obviously, then, this spinodal must lie below the experiment line of homogeneous ice nucleation temperatures, but not by much. The analysis of diverging compressibility by Kanno and one of us, indicated that the gap $\Delta$T to be about 5K. It may be accessible, near the high pressure end of the ice I homogeneous nucleation temperature T$_h$, by the type of spinodal decomposition study performed recently by Murata and Tanaka\cite{murata2012liquid}, but would need the development of a faster analysis. We note that it is this line that Holten and Anisimov\cite{holten2012entropy} have recently adopted as their coexistence line in an apparently successful isoenergetic two state treatment of the water problem.  

This modified EoS scheme with a liquid-liquid coexistence line preceding the HDL spinodal line and extending far into the negative pressure domain, has been designated the ``critical point-free" scenario\cite{angell2008insights}. It represents an extreme case of the liquid-liquid scenario, in which the critical point has been pushed all the way to the liquid-vapor spinodal and has become inextricably merged with it. Intermediate cases where the LLCP lies at negative pressure, but falls short of the liquid-vapor spinodal, would be difficult to distinguish from the critical point-free scenario. Both require that the experimental data acquired in ``no man's land" at either positive or negative pressure, will terminate during temperature decrease at a first order transition. This must occur either at the coexistence line, or at the underlying spinodal. At negative pressure this would release the tension and would therefore be a stabilizing event.
 
\subsection{Hard data in the negative pressure domain}
What hard data do we have to work with in the negative pressure domain?  As noted above, stretched water measurements are sparse in the literature. A good review has been given by Caupin et al. recently\cite{caupin2012exploring}. The majority of earlier experiments such as the Speedy-Henderson spiral capillary Bourdon tube\cite{henderson1980berthelot}, and the Winnick-Cho centrifugal stretcher\cite{winnick1971pvt} have only been able to reach about -20 MPa,  while recent microfluidic transpiration across a membrane\cite{wheeler2008transpiration} and the acoustic methods\cite{davitt2010water,caupin2012exploring} have not done much better despite their increased sophistication. These are far short of the tensions reached by the alternative micro-Berthellot tube methods described below, or the micro-capillary rupture studies of Machin and coworkers\cite{machin2002limiting}. The existence of a second, somewhat mysterious, ``no-man's land" has been suggested (Fr\'ed\'eric Caupin, private communication).

The most recent and most sophisticated effort, which is unlikely to be bettered by current techniques, is that by Davitt et al.\cite{davitt2010water,azouzi2012coherent} who made direct measurement of the pressure at a low density rarefaction created by a high frequency sound wave, the hypersonic velocity in the rarefaction zone being determined by Brillouin scattering of a laser beam. The results confirm the accuracy to -26 MPa of both HGK and IAP-WS95 equations which track each other perfectly to this limit, and only separate significantly beyond -100 MPa.  This point is identified in Figure 7.

\begin{figure}[ht]
\centering
\includegraphics[scale=0.1]{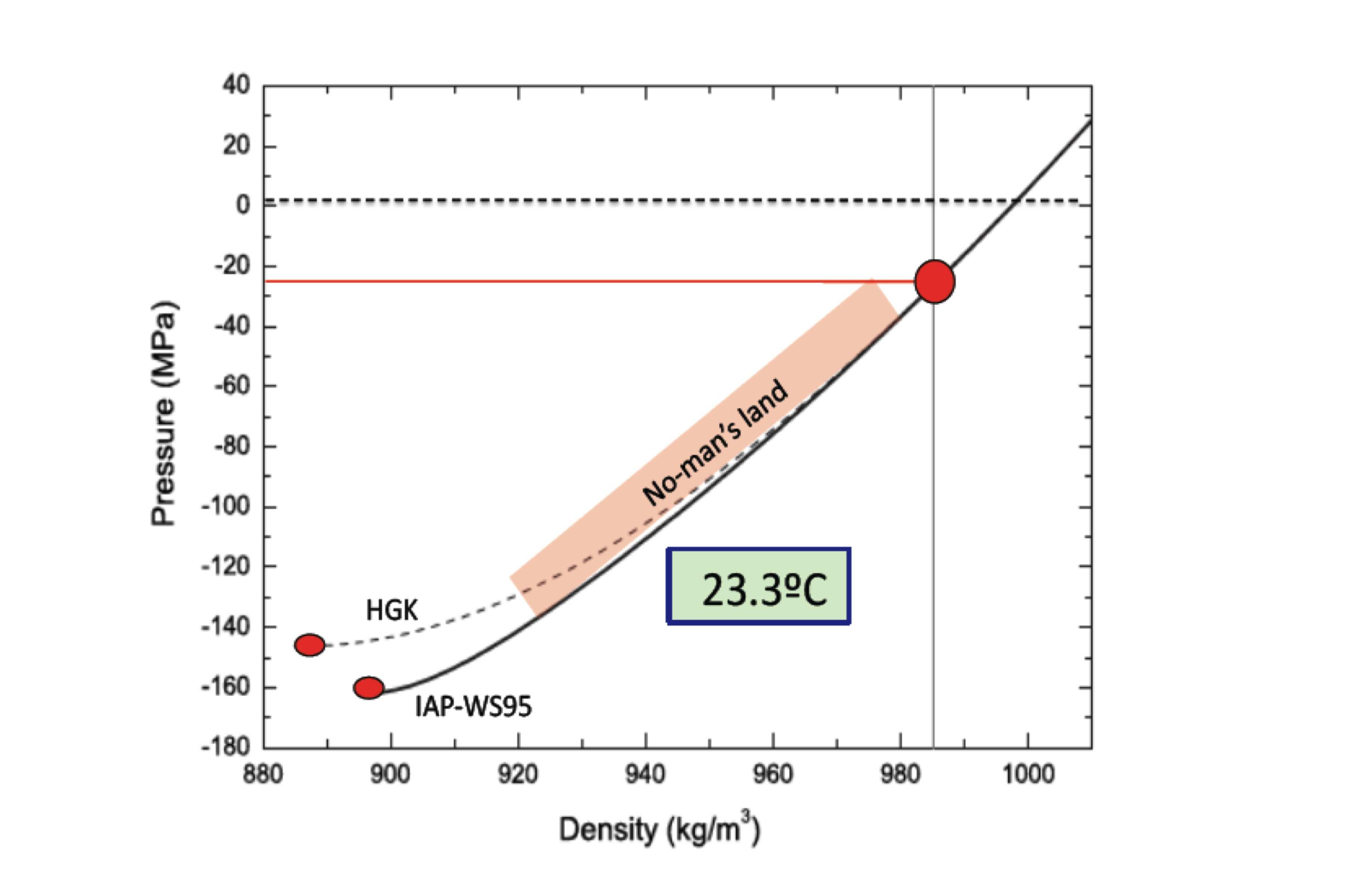}
\caption{The directly measured isotropic tension (negative pressure) sustained by ultraclean water of density 925 kg/m3, (large solid point) from refs\cite{davitt2010water,machin2002limiting}. The value is exactly that predicted by extrapolations of both of HGK and IAP-WS95 equations of state. The smaller red points are maximum tensions (spinodal limits) reached by water  at 23.3$^\circ$C, according to these EoSs. Note the diverging compressibilities as the limit is approached.}
\end{figure}

Alternative to these limited studies are the microscopic Berthelot tube isochoric  studies of the limiting tension of water (and its solutions) at different densities. These are created by autoclaving cracked mineral crystals in the presence of water, to lock in 1-15 $\mu$m diameter isolated inclusions that are filled with water, at the temperature and pressure of autoclaving, as the cracks heal\cite{Zheng1991}.  On removal from the autoclave and cooling, the filled vesicles exceed their tensile limits and cavitate. To perform an experiment the quartz blocks are heated again under microscope observation to determine the temperature where the vapor bubbles in  each vesicle disappear. This fixes the isochore density by reference to the precisely known equation of state. They are then cooled under observation until cavitation is observed and the temperature noted.

Since the cavitation depends on a nucleation process that is random in space and time, and may also be heterogeneously initiated, multiple determinations must be made and the lowest temperature observed taken as the homogeneous nucleation temperature for the sample density. Since the natural criticism of such a method is that it will be controlled by impurities dissolved from the host matrix, the data of Zheng et al.\cite{zheng2002limiting}, 
shown in Figure 8, were obtained using three different mineral hosts, each of which provided its own characteristically shaped vesicles. The fact that little if any difference can be seen between the different hosts, within the natural scatter, tells us that there is no matrix dependence of the results. The data have recently been confirmed by detailed studies using quartz matrices reported by Shmulovich et al.\cite{shmulovich2009experimental} that we will discuss in more detail elsewhere.

\begin{figure}[ht]
\centering
\includegraphics[scale=0.1]{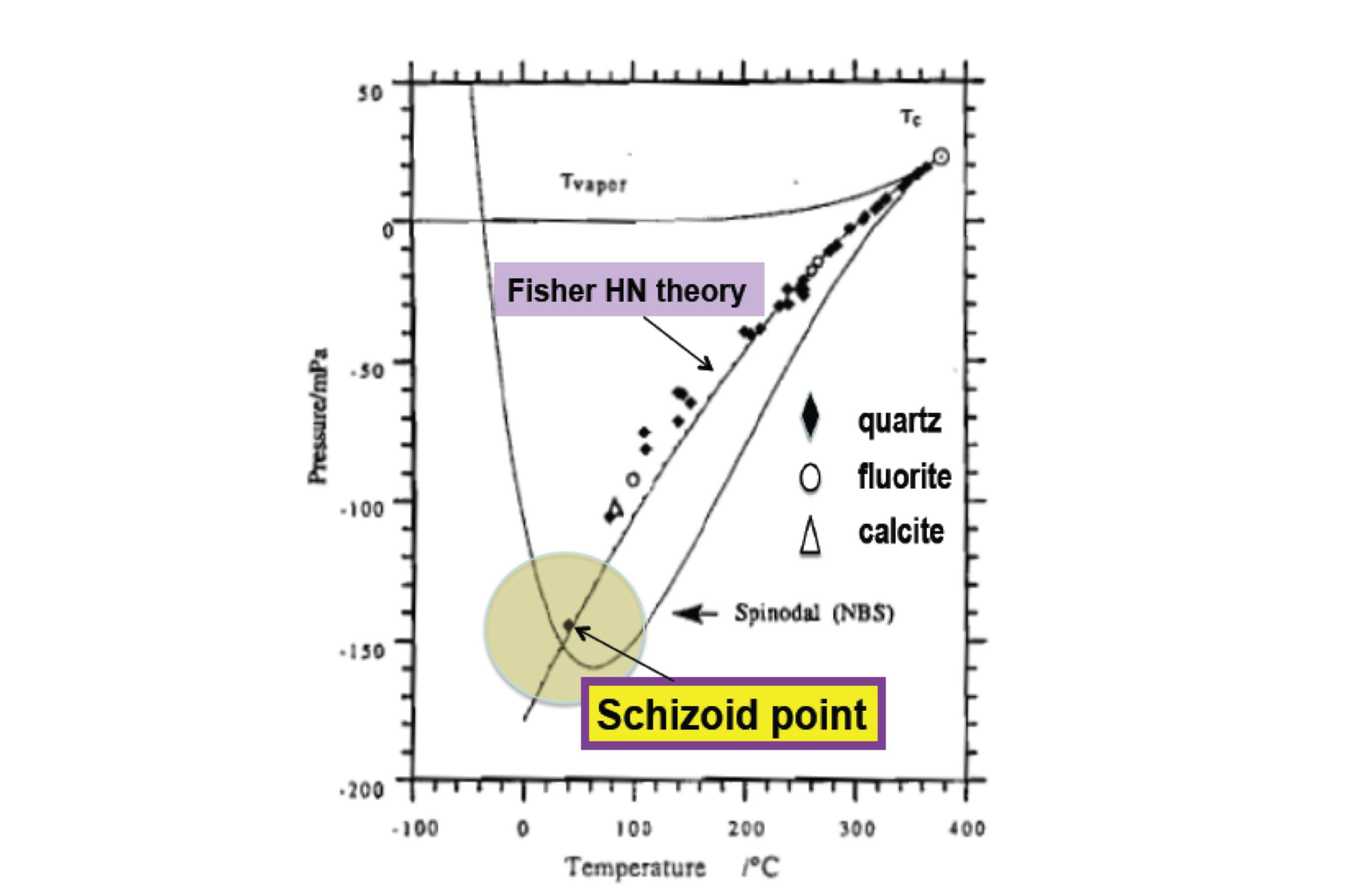}
\caption{Lowest temperatures of cavitation of pure water inclusions contained in alternative mineral matrices, quartz, fluorite and calcite. The lowest temperature inclusion result (designated ``schizoid point") is one of three samples that cavitated unpredictably (between 40 and 47$^\circ$C) in some of multiple repeat cooling runs, and sometimes did not cavitate at all, showing that they all reached, or passed through, a point of maximum tension (where nucleation was strongly time-dependent) This unambiguously identifies a TMD at large negative pressure. The exact pressure at this TMD will depend on the EoS used to extrapolate the isochore. Questions raised about the accuracy of observation due to vesicle shape have been put to rest by the agreement with more recent data from Shmulovich et al.\cite{shmulovich2009experimental}}
\end{figure}

The unpredictable cavitation at the lowest temperature point, and the total absence of cavitation in some runs, identifies this isochore as having a tensile stability limit that coincides with the temperature of maximum density. The TMD so identified is so close to the spinodal limit according to either of the HGK or IAP-WS95 EoS that any reversal to avoid a merging of LMD with the spinodal seems unlikely. The actual pressure is dependent on the equation of state used to predict the isochore, but the temperature of the TMD is unambiguous. 

More recent studies focused on the nucleation mechanism have been reported by El Mekki Azouzi and coworkers\cite{azouzi2012coherent}, concluding that the cavitation near the density maximum is indeed well described by classical nucleation theory subject to minor and physically reasonable adjustments of the surface tension used in the calculations.

The availability of stretched water samples of multimicron dimensions has opened the door for direct physical property measurements of water under high tension, under conditions of constant volume (the default choice for theoretical studies).  Alvarenga et al.\cite{alvarenga1993elastic} were the first to take advantage of this opportunity. They published a detailed study of the Brillouin scattering from water over the range 0-400$^\circ$C, determining the sound velocity, hence the adiabatic compressibility, reaching pressures as negative as -100 MPa before cavitation terminated the measurements. Unlike the sound velocity in water in coexistence with its vapor, the temperature dependence at constant volume is always positive, but evidently passes through a minimum somewhere in the temperature range 50-150$^\circ$C. These authors gave independent arguments for the validity of equation of state estimates of the tension.

Extensions of these measurement have now been made by Caupin and coworkers (to be published) but are yet unpublished so cannot be discussed at this time. The presence of a density maximum should be detectable, in the absence of cavitation (i.e. for isochore densities less than 0.93gml$^{-1}$, by the observation of a minimum (ideally a zero) in the Landau-Placzek ratio, which theory requires whenever C$_p$ = C$_v$, e.g. at a density maximum. The Landau-Placzek ratio is not mentioned in ref.\cite{alvarenga1993elastic}. In view of the foregoing discussion and Figure 3, the study of the Landau-Placzek ratio in non-cavitating samples of densities between 0.9 and 1.0 would seem important.  Also to be mentioned among the direct characterization studies of water under tension are the Raman spectra.  These were first studied (by Green et al.)\cite{green1987coupling} on water in coiled silica spirals in which the pressure could be measured directly. More recent Raman spectra have re-explored higher tensions in inclusion samples but at the cost of direct pressure knowledge. Investigation of samples with densities \textless 0.91 are currently the highest priority.

\subsection{A modified van der Waals model for stretched liquids}

\begin{figure}[ht]
\centering
\includegraphics[scale=0.12]{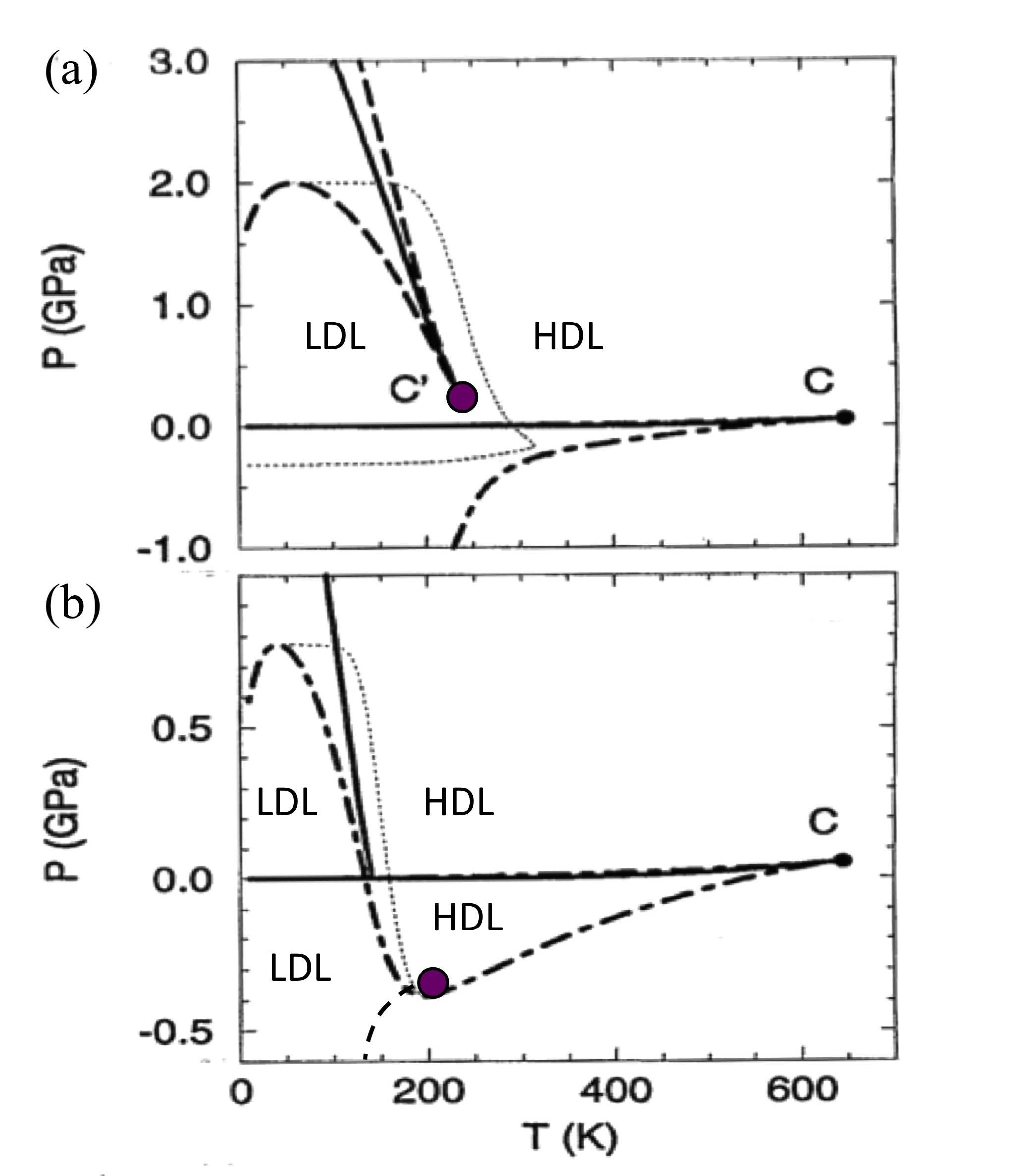}
\caption{Phase diagrams for the Poole model (a) strong H-bond case, with second critical point at positive pressures. (b) Weak H-bond case where  the second critical point merges with the liquid vapor spinodal and only the liquid-liquid coexistence line with its spinodal limit to the high density liquid phase, remains. The dashed line addition to the original figure of Ref.\cite{poole1994effect} recognizes that the LDL-phase should have its own spinodal limit.}
\end{figure}

Helpful in the search for understanding of  these quite complex relationships is a modification of  the original van der Waals equation, due to Peter Poole\cite{poole1994effect}. Intended originally as a means for understanding the generation of second critical point in water, it also produces the ``critical point-free" scenario, by modification of a single parameter in the model. Poole's simple concept is that, if a state of \textit{low} energy, and volume \textit{intermediate} between the gas density and the closepacked liquid density characteristic of simple molecular liquids, can exist, then the coexistence domain (familar in the temperature vs density plot for simple liquids), will be split in two parts. The low density part will exhibit a ``normal" critical point where the densities of intermediate liquid and gas become the same, while the high density part will be characterized by the coexistence of two liquids of different density which, in their turn, can become equal at a liquid-liquid critical point. Motivated by water, the intermediate density liquid was given an energy determined by the strength of hydrogen bonds, generating an open network random ice-like network structure. A parameter describing the width of the gaussian for the volume of the intermediate density phase completed the parameterisation of the model. It was then formulated to generate an equation of state, with consequent phase diagrams. Two of these, the upper for ``strong" bonds (or strong tetrahedrality) and the lower for ``weak bonds" (insufficient to truly split the coexistence curve) are shown in figure 9. The latter is the one that is most consistent with the empirical EoS, and for the findings we have presented for behavior in the negative pressure domain. However, the behavior of the density maximum in the ``strong" bond case (Figure 9 upper panel) shows that models with high pressure (\textgreater0.2 GPa) second critical points (TIP5P and TIP4P2005) may not be excluded solely on the basis of the density maximum observed at negative pressures (Figure 8). While direct calculations of the density maximum behavior, like those illustrated in Figure 3, would seem to exclude the TIP4P model (and any other with the same behavior), there are other considerations to sustain the view that water models with critical points close to ambient pressure should prove the most satisfactory\cite{mishima2010volume,holten2012entropy}. 

Other models like that of Franseze et al.\cite{franzese2003intramolecular,stokely2010effect}, based on lattice statistics with cooperativity, that of Tu et al.\cite{tu2012different}, based also on energy cooperativity, and that of Stillinger and Weber\cite{stillinger1985computer}, based on repulsive discouragement of low tetrahedrality, exist but are not discussed here. We believe that the main challenge now is to develop new methods for making physical measurements at negative pressure. NMR and molecular ion probe spectroscopy of dilute impurity signalling species, like HCN and HN$_3$, might offer good opportunities.

\section{Concluding commentary}
In spite of the intensity with which water has been studied, it continues to guard its secrets closely. Ideas about water's behavior have been built around its hydrogen bonding behavior and its many anomalies are attributed to the unique bonding that its ``two hands, two feet"\cite{ball2001life} character bestow. Many of these anomalies in the supercooled state have pointed to a singularity existing beyond the homogeneous crystallization temperature. Novel experiments probing the metastable stretched water regime are seen as promising ``scenic routes" for reaching a deeper understanding of this remarkable liquid. Its continuing mysteries hopefully provide not only challenges but inspiration for many future studies of liquid water behavior.

\section*{Acknowledgement}
The authors acknowledge the support of the Italian Physical Society during the Enrico Fermi School on water, at which their writing partnership was forged. CAA acknowledges support of the NSF under Grant no. CHE 12-13265. CAA appreciates helpful discussions with Sergey Buldyrev and Paola Gallo, and the permission to use the original superposition of shifted TIP4P data and experimental data from Ref.18, and its overlay with the locus of TMDs calculated from the IAP-WS95 EoS, in our Figure 3 plot comparing TIP4P with equation of state (IAP-WS95) predictions for the loci of TMD with pressure. The plot was generated using the program developed by Stephen L. Moshier.

\bibliographystyle{unsrt}
\end{document}